\documentclass[mathleft, final]{an}
\usepackage{graphicx}
\usepackage{times}
\def\PN{{\cal P\!N\!}}   

\begin{document}
\Pagespan{1}{4}           %
\Yearpublication{2008}  %
\Yearsubmission{2008}   %
\Month{}                %
\Volume{}               %
\Issue{}                %
\title{Post-Newtonian simulations of super-massive black hole
       binaries in galactic nuclei}
\author {I.~Berentzen\inst{1}\fnmsep\thanks{Corresponding author:
         \email{iberent@ari.uni-heidelberg.de}\newline},
         M.~Preto\inst{1},
         P.~Berczik\inst{1,2},
         D.~Merritt\inst{3}, \and
         R.~Spurzem\inst{1}
}
\titlerunning{Post-Newtonian simulations of SMBH binaries}
\authorrunning{I.~Berentzen et al.}
\institute{
 Astronomisches Rechen-Institut, Zentrum f\"ur Astronomie, Universit\"at
 Heidelberg, M\"onchhofstr. 12--14, D-69120 Heidelberg, Germany
\and 
 Main Astronomical Observatory, National Academy of Sciences
 of Ukraine, 27 Akademika Zabolotnoho St., 03680 Kyiv, Ukraine
\and 
 Center for Computational Relativity and Gravitation, Rochester
 Institute of Technology, 78 Lomb Memorial Drive, Rochester, NY 14623
}
\received{} \accepted{}
\publonline{later}
\abstract{%
 We present some preliminary results from recent numerical simulations
 that model the evolution of super-massive black hole (SMBH) binaries
 in galactic nuclei. Including the post-Newtonian terms for the binary
 system and adopting appropriate models for the galaxies allows us, for
 the first time, to follow the evolution of SMBH binaries from kpc
 scales down to the coalescence phase. We use our results to make
 predictions of the detectability of such events with the gravitational
 wave detector LISA\@.%
}
\maketitle
\section{Introduction}
 The formation and evolution of super-massive black holes (SMBHs)
 in galactic nuclei is an important ingredient for our understanding of
 galaxy formation and evolution. In the hierarchical cosmological framework,
 galaxies merge in the course of their evolution. Therefore, either many
 massive binary (or even multiple) SMBHs should exist, or else we need
 to develop a detailed understanding of what ultimately happens to the
 massive binaries.
 According to the standard theory, the evolution of SMBHs after
 a galaxy merger can be divided in three stages (Begelman, Blandford
 \& Rees 1980): {\bf (i)} Dynamical friction causes the individual
 SMBHs to spiral to the galactic centre where they form a binary.
 {\bf (ii)} Super-elastic scattering of stars causes the binary to harden.
 {\bf (iii)} Eventually, the binary may coalesce owing to the
 emission of gravitational waves. However, this is only possible if stellar-
 and gas-dynamical processes have first brought the two SMBHs to small enough
 separations (some $10^{-3}$\,pc) that the emission of gravitational waves
 is significant.\ Wheth\-er Nature typically succeeds in
 overcoming this ``final parsec problem,'' (see review by
 Milosavljevi{\'c} \& Merritt 2003) is currently unknown.

 Here we present preliminary results of recent $N$-body simulations of
 SMBH binaries in galactic nuclei, including post-Newtonian ($\PN$)
 equations of motion up to the 2.5$\PN$ order. In Sec.~\ref{toy}
 we describe our $\PN$ implementation and provide numerical tests
 and simple two-body experiments. In Sec.~\ref{SMBH} we summarise some
 first results of our detailed galactic nuclei simulations, following
 the evolution of the SMBH binaries from the unbound state down to 
 relativistic coalescence. We conclude with Sec.~\ref{Concl}.

\section{A two-body Hermite ${\PN}\,$ integrator}           \label{toy}
 In order to incorporate the dynamics of relativistic binary systems
 into our direct $N$-body code (see Sec.~\ref{SMBH}) we use the
 $\PN$ equations of motion, which are expressed by expanding
 the relativistic acceleration between two compact objects in a power
 series of $1/{c}$ (e.g., Damour \& Deruelle 1981; Soffel 1989).
 Schematically, the ${\PN}$ equation of motion for an object in the
 binary can be written as:
\begin{equation}
\label{eq1}
 {\bf a} = {\bf a}_0
      + \frac{1}{{c}^2} {\bf a}_1
      + \frac{1}{{c}^4} {\bf a}_2
      + \frac{1}{{c}^5} {\bf a}_{2.5}
      + {\cal O}\left( \frac{1}{c^6} \right),
\end{equation}
 \noindent
 where ${\bf a}_0$ is the {\em classical} Newtonian acceleration,
 ${\bf a}_1$ and ${\bf a}_2$ are the energy conserving $1\PN$ and
 $2\PN$ corrections, respectively, and ${\bf a}_{2.5}$ is the
 dissipative $2.5\PN$ term, describing the dominant damping forces
 due to gravitational wave emission.
 In the current $\PN$ routine we neglect any effects of spin-orbit
 and spin-spin coupling (e.g., Damour 1987). For the full expressions
 of the individual terms up to 3.5$\PN$ order see, e.g., Blanchet (2006).
 We implement the equations of motion formulated in a general harmonic (Cartesian)
 coordinate system, whereas other astrophysical implementations often use
 expressions given in the binary's centre of mass frame
 (e.g., Kupi, Amaro-Seoane \& Spurzem 2006; Aarseth 2007).

\begin{figure}[tpb]
\begin{center}
  \includegraphics[width=1.0\columnwidth]{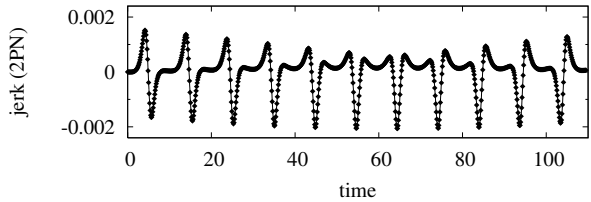}  
\end{center}
\caption{Comparison of the $\PN$ implementation (solid line)
 and the finite differences $\Delta a_i/\Delta t_i$ (symbols). Shown
 are the curves for the $2\PN$ x-component of some
 (arbitrary) orbit integration.}
\label{fig1}
\end{figure}

\begin{figure}[tpb]
\begin{center}
  \includegraphics[width=0.7\columnwidth]{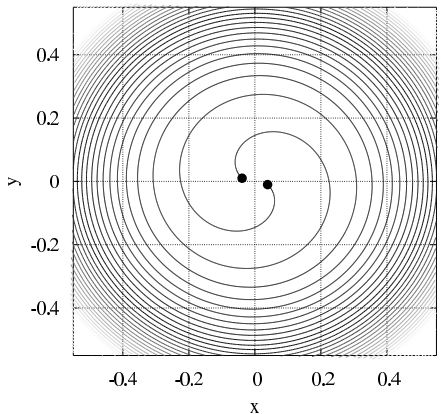}  
\end{center}
\caption{Inspiral and plunge of an equal-mass binary system using the
 post-Newtonian equations of motion. The numerical integration is done
 using a 4$^{th}$ order Hermite integrator and hierarchical time-step
 scheme.} 
\label{fig2}
\end{figure}

 In order to achieve high accuracy in the simulations we use a $4^{th}$
 order Hermite integrator (e.g., Makino \& Aarseth 1992), which requires
 in addition to the accelerations also their first time-derivatives
 (``{\em jerk}'') -- including the $\PN$
 corrections. Since the $\PN$ accelerations and jerks are
 complicated functions of the particles' masses, positions and velocities,
 their derivation and numerical implementation is a potential source
 of error. Therefore, we provide some numerical tests which turned
 out to be useful benchmarks, as already applied with other
 $\PN$ implementations (e.g., Kupi et al.~2006; L\"ockmann
 \& Baumgardt 2008):

 {\bf (i) Time-derivatives.} To check the correct implementation of
 the jerks of the $\PN$ terms, we evolve a binary system for several
 orbital periods and compare the finite differences
 $\Delta {\bf a}_{\PN}/\Delta t$ for consecutive time-steps to
 the directly calculated $\dot {\bf a}$ from our implementation
 (Fig.~\ref{fig1}). Any deviations between the two would be
 easily detectable and clearly indicate any errors
 in the values of the computed jerk.
 In Fig.~\ref{fig2} we show a typical trajectory of an binary inspiral
 integrated with our ${\PN}$ code, starting with a quasi-circular
 orbit. 
 
   
 {\bf (ii) Merging time for eccentric binary orbits.} In the following
 test we set the gravitational constant $G$ and the two particle
 masses to unity. The speed of light in our model units is set to
 ${c}\!=\!10$.\footnote{These units are chosen for numerical testing
 purposes only and are not physically motivated.}
 The two particles are placed on the $x$-axis with an initial separation
 of $\Delta x\!=\!2$. The initial velocities $\pm v_y$ are calculated for
 a Keplerian orbit of given eccentricity $e_0$. We then integrate the
 orbits using {\bf (a)} the {\em full} $\PN$ (i.e., in our case up to 2.5$\PN$~)
 corrections and {\bf (b)} applying only the 2.5$\PN$ correction. In both
 cases the integrations are stopped at the times $T_{\PN}$ and $T_{2.5\PN}$,
 respectively, when the binaries reach a separation of $10$ Schwarzschild radii
 ($R_{\mathrm S}\!=\!2{G}m_{\mathrm{bh}}/{c}^{2}$), i.e., well
 before the ${\PN}$ approximation may become inadequate.
 The results for different initial eccentricities are given in Table~\ref{tab1}.
 The numbers given in Table~\ref{tab1} are in agreement with integrations using
 the $\PN$ implementations of Kupi et al.~(2006) and Aarseth (2007). The exact
 result may slightly vary for different integration schemes and time-step controls.

\begin{figure}[tpb]
 \begin{center}
  \includegraphics[width=1.0\columnwidth, height=5.5cm]{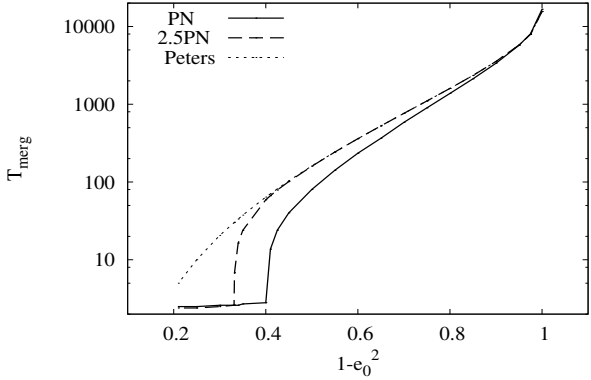}        
 \end{center}
 \caption{Shown are the results obtained from numerical integration of the
 orbit-averaged equations from  Peters (1964) (dotted line) and  from the
 direct two-body integration including the $2.5\PN$ correction (dashed line)
 and full $\PN$ corrections.}
 \label{fig3}
\end{figure}

\begin{table}
\begin{center}
\caption{The columns from left to right give for each integrated orbit
 the corresponding  $(1\!-\!e_0^2)$, the initial eccentricity $e_0$, the
 initial semi-major axis $a_0$, and the two 'merging' times
 $T_{\PN}$ and $T_{2.5\PN}$, respectively.}
\label{tab1}
\begin{tabular}{ c c c r r }
\hline
 $1-e_0^2$ & $e_0$ & $a_0$ & T$_{\cal PN}$ & T$_{2.5\,\cal{PN}}$ \\
\hline
 1.000  &  0.000 &  2.000 & 16666.2 & 15623.4 \\
 0.950  &  0.224 &  1.635 &  5741.4 &  5783.1 \\
 0.900  &  0.316 &  1.519 &  3370.4 &  3554.0 \\
 0.850  &  0.387 &  1.442 &  2125.9 &  2347.1 \\
 0.800  &  0.447 &  1.382 &  1377.7 &  1598.0 \\
 0.750  &  0.500 &  1.333 &   899.7 &  1102.7 \\
 0.700  &  0.548 &  1.292 &   585.3 &   763.7 \\
 0.650  &  0.592 &  1.257 &   365.6 &   527.1 \\
 0.600  &  0.632 &  1.225 &   234.6 &   360.5 \\
 0.550  &  0.671 &  1.197 &   141.7 &   242.9 \\
 0.500  &  0.707 &  1.172 &    80.3 &   159.7 \\
\hline
\end{tabular}
\end{center}
\end{table}
 In Fig.~\ref{fig3} we show a comparison of the {\em ``merging''}
 times $T_{\PN}$ (solid line) and $T_{2.5\PN}$ (dashed line). Note
 that the results can vary by a factor of a few with increasing
 eccentricity. This underlines the importance of the lower order
 $\PN$ corrections, which sometimes have been neglected in earlier
 astrophysical simulations.
 The pericentre shift due to the $1\PN$ and $2\PN$ terms leads to
 more pericentre passages and therefore to an overall stronger
 dissipation (especially for high eccentricities), which
 shortens $T_{\PN}$ as compared to $T_{2.5\PN}$.
 The almost horizontal part of the two
 curves in Fig.~\ref{fig3} ($1-e^2_0\!\approx\!0.2\!\ldots\!0.4$)
 is the result of a direct plunge into the $10\,R_{\mathrm S}$ region
 due to the high eccentricity and small pericentre distances.
 Finally, we compare our results  to the numerical integration of
 the equations given by Peters (1964), describing the orbit-averaged
 changes of the eccentricity and semi-major axis due to gravitational
 wave emission. We find good agreement between the latter integration
 and our corresponding 2.5${\cal P\!N}$ two-body integrations. The
 diverging behaviour for $(1\!-\!e_0^2)\!<\!0.4$ comes from the
 fact that our orbit-averaged integration is terminated when the semi-major
 axis (instead of the binary separation) reaches $10\,R_{\mathrm S}$.

\begin{figure}[tpb]
\begin{center}
  \includegraphics[width=1.0\columnwidth, height=5.5cm]{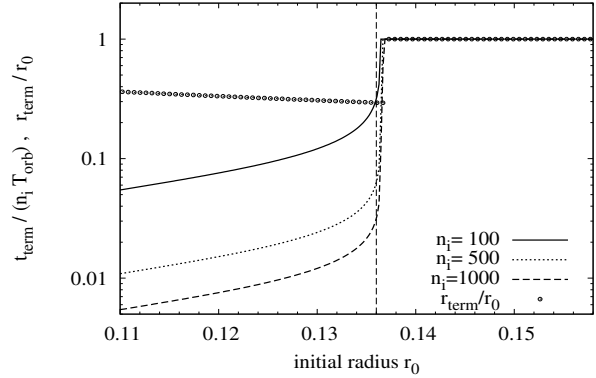}  
\end{center}
\caption{Normalized integration time $t_{\mathrm{term}}/(n_i \times T_{\mathrm{orb}})$
 for initially circular orbits with radius $r_0$. We show our results for
 $n_i\!=\!100$ (solid line), $n_i\!=\!500$ (dotted line) and $n_i\!=\!1000$
 (dashed line). The symbols show the final ratio $r_{term}/r_0$. The vertical line
 indicates the radius of the innermost stable circular orbit (ISCO) as
 derived from the linear stability analysis. Note that our $\PN$ code is capable of accurately
 capturing the ISCO.}
\label{fig4}
\end{figure}

 {\bf (iii) The innermost stable circular orbit (ISCO).} Here we
 study the linear stability of circular orbits. It is known for
 binary systems that circular orbits with radii below a certain
 radius $r_{\mathrm I}$ are unstable, even without dissipative 
 effects. The evolution of an relativistic
 binary changes from a quasi-adiabatic inspiral to an unstable plunge
 when crossing $r_{\mathrm I}$ (e.g., Kidder, Will \& Wiseman 1993).
 Analytically, the radius $r_{\mathrm I}$ of the ISCO can be
 calculated using the standard linear stability analysis (e.g.,
 Blanchet \& Iyer 2003). To determine $r_{\mathrm I}$ on an ISCO
 for objects of comparable mass we use here the exact 2$\PN$ accurate
 circular frequencies $\omega_{2\PN}$ as given in Preto (2007). We find
 for the binary system described in test (ii) that the ISCO has a
 radius of $r_{\mathrm I}\!\sim\!0.136$ in our model units (Preto 2007).
 Now we determine $r_{\mathrm I}$  by numerical integration of circular
 $\PN$ orbits. The model units and masses are the same as the ones used
 in test (ii). The initial velocities of the two particles are calculated
 based on the 2$\PN$ accurate $\omega_{2\PN}$ (see above).
 We integrate the $\PN$ circular orbits including only $1\PN$ and
 $2{\cal P\!N}$ corrections, i.e., under the absence of gravitational
 wave emission. The integration of the orbits has been terminated
 at a time $t_{\mathrm{term}}$ either after an unstable plunge or
 after a total maximum of $n_i$ (initial) orbital periods $T_{\mathrm{orb}}$. 
 In Fig.~\ref{fig4} we show $t_{\mathrm{term}}/(n_i\!\times\!T_{\mathrm{orb}})$
 for $n_i\!=\!100$ (solid line), $n_i\!=\!500$ (dotted line) and 
 $n_i\!=\!1000$ (dashed line). As an indicator for stability we also
 plot the final ratio $r_{\mathrm{term}}/r_0$, where $r_{\mathrm{term}}$ is the
 separation of the particles at time $t_{\mathrm{term}}$.
 We find that our $\PN$ two-body integration accurately captures the
 radius of the ISCO in all three cases. The transition from the stable
 to the unstable regime is very abrupt as the theory predicts, and
 happens at the radius estimated by the linear stability analysis.
%
\section{Super-massive black hole binaries in galactic nuclei} \label{SMBH}
 In this section we present preliminary results of recent direct $N$-body
 simulations of SMBH binaries in galactic nuclei including $\PN$
 corrections (Berentzen et al.~2008). Initial conditions were based
 on the models of Berczik et al.~(2006). The initial field particle
 distribution was a discrete realisation of the axisymmetric, rotating
 King (1966) models of Longaretti \& Lagoute (1996).
 We consider such models to be reasonable representations of galactic nuclei
 in post-merger galaxies.
 When the rotation parameter is sufficiently large, the models evolve
 rapidly into triaxial bars. Berczik et al.~(2006) found that in this
 case, the massive binary continued to harden at an essentially
 $N$-independent rate, due presumably to the centrophilic nature
 of the orbits (Merritt \& Poon 2004). This feature
 means that evolution of the massive binary can be faithfully reproduced
 even with modest particle numbers. Accordingly, the 
 nucleus was represented here using either $N\!=\!25\!\times\!10^3$ or
 $N\!=\!50\!\times\!10^3$ particles, in each case with nine different
 random realisations. The masses of the two SMBH particles were
 chosen to be one percent of the total mass. For the simulations we used
 an updated version of the publically
 available\footnote{http://wiki.cs.rit.edu/view/GRAPEcluster/phiGRAPE}
 $\varphi$\,-{\sc Grape} code (Harfst et al.~2007). The code is an
 {\sc Nbody1}-like algorithm (Aarseth 1999), including a hierarchical
 time-step scheme and a $4^{th}$-order Hermite integrator. Our modified
 version includes the $\PN$ implementation described in
 the previous section. Our simulations were carried out on the
 high-performance {\sc Grape}-6A clusters at the Astronomisches
 Rechen-Institut (Heidelberg), Rochester Institute of Technology (New York) 
 and the Main Astron. Observatory (Kiev). More details on the code and
 hardware are given in Berentzen et al.~(2008) and Spurzem et al.~(2008).

\begin{figure}[tpb]
\begin{center}
  \includegraphics[width=1.0\columnwidth, height=4.8cm]{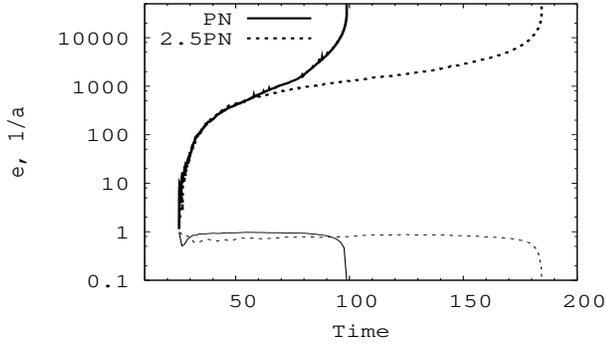}  
\end{center}
\caption{Evolution of the SMBH binaries in a model of a rotating
 galactic nucleus. Shown are the orbital eccentricity $e$ (thin lines) and the
 inverse semi-major axis (thick lines) as a function of time. The full
 and dashed lines represent models using the full $\PN$ and $2.5\PN$
 corrections, respectively. Note the significantly different evolution
 of the binary.}
\label{fig5}
\end{figure}

\begin{figure}[tpb]
\begin{center}
  \includegraphics[width=1.0\columnwidth, height=5.5cm]{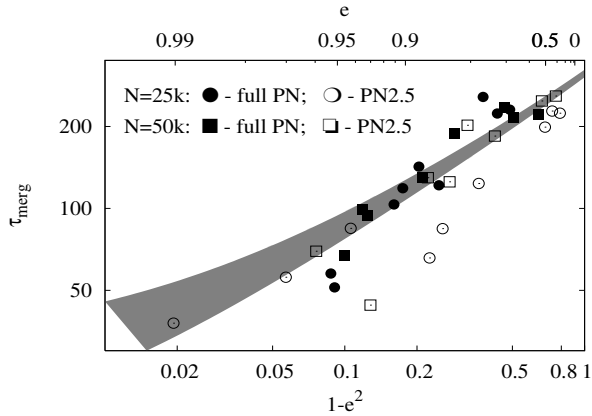}  
\end{center}
\caption{Merging time $\tau_{\rm merg}$ as a function of eccentricity $e$.
 Shown are simulations with 25k (circles) and 50k (squares) particles.
 The open and filled symbols represent models with full $\PN$ and only
 $2.5\PN$ corrections, respectively. The gray region marks the theoretically
 estimated merging times (see Berentzen et al.~2008).}
\label{fig6}
\end{figure}

 In Fig.~\ref{fig5} we show the evolution of the inverse semi-major axis 
 $1/a$ and eccentricity $e$ of the binaries in our simulations with {\em full}
 $\PN$ (solid lines) and with only 2.5$\PN$ corrections (dashed
 lines). Initially the two SMBH particles are unbound; they form a very
 eccentric binary at $t\!\sim\!25$. Afterwards the binaries steadily
 harden mainly by stellar-dynamical processes, before relativistic
 effects start to dominate their evolution. Eventually, they reach the
 inspiral pha\-se during which the orbits circularise before the final coalescence.
 As expected, we find that the lower order $\PN$ terms significantly
 influence the evolution of the binaries and must not be neglected in 
 this type of simulation. Finally,  we show in Fig.~\ref{fig6} the merging
 times $\tau_{\mathrm{merg}}$ for our SMBH binaries as a function of $e$.
 Scaling our model units to real galaxies we find that $\tau_{\mathrm{merg}}$
 in our models is typically less than $1$~Gyr after the binary has formed. 
\section{Summary} \label{Concl}
 We upgraded the direct $N$-body code $\varphi$-{\sc Grape} by including
 a relativistic treatment for binary SMBHs via the $\PN$ equations
 of motion up to 2.5$\PN$ order. The new implementation was carefully
 tested for accuracy. We then carried out numerical simulations of SMBH
 binaries in galactic nuclei. Our models allow us, for the first time,
 to follow the evolution of initially unbound SMBHs with kiloparsec-scale
 separations down to relativistic
 inspiral and coalescence. The SMBH binaries in our simulations form generally
 with high eccentricities, which are maintained during the evolutionary phase
 dominated by stellar-dynamical effects. We have shown that the inclusion of
 the lower order $1\PN$ and $2\PN$ corrections is crucial for obtaining the
 correct time dependence of the binary orbital parameters and 
 final coalescence time. Our results
 demonstrate that SMBH binaries in certain models of galactic nuclei can
 overcome the stalling barrier by stellar-dynamical effects alone.
 A detailed description and discussion of our models and results will be
 given in Berentzen et al.~(2008) and in Preto et al.~(2008).
\acknowledgements
 We cordially would like to thank S. Aarseth, P. Amaro-Seoane, A. Gopakumar,
 A. Gualandris, S. Harfst, G. Kupi and G. Sch\"a\-fer for valuable
 suggestions and fruitful discussions. This work was supported by Volkswagen
 Foundation ({\sc Gra\-ce}, Ref. I/80\,041-043), DLR (Deut\-sches Zentrum
 f\"ur Luft- und Raumfahrt), and SFB 439 of DFG\@.
 I.B. thanks the conference organisers for financial support.

\end{document}